\documentclass[12pt]{article}
\input{epsf.tex}

\begin{document}

\title{The Auxiliary Field Method as a Powerful Tool \\ for
Nonperturbative Study}
\author{Taro Kashiwa\thanks{taro1scp@mbox.nc.kyushu-u.ac.jp}
\\
Department of Physics, Kyushu University, Fukuoka 812-8581,
JAPAN}
\maketitle
%
%
%
\begin{abstract}
The auxiliary field method, defined through introducing an
auxiliary
 (also called as the Hubbard-Stratonovich or the Mean-) field
 and utilizing a
loop-expansion, gives an excellent result for a
wide range of a coupling constant. The analysis is made for
Anharmonic-Oscillator and Double-Well examples in 0-(a simple
integral) and
1-(quantum mechanics)dimension. It is shown that the result
becomes more
and more accurate by taking a higher loop into account in a
weak coupling region,
however,
it is not the case in a strong coupling region. The 2-loop
approximation is
 shown to be still
insufficient for the Double-Well case in quantum mechanics.

\end{abstract}
%


\section{Introduction}
In most of actual situations, path integral expressions are
given as a non-gaussian form so that
some approximation is always needed. Apart from perturbative
treatment, such as
a weak(strong) coupling expansion, which can only describe a
small (large) coupling
region, we have sought for other recipes to be able to handle
cases for a wider
coupling range: the variational method\cite{FH} has been well
known and applied
successfully to the polaron problem\cite{SS}. The method is
combined with an
optimization technique and has been actively
discussed\cite{VOP}. Numerical
estimation is also possible once expressed in the path integral
form: for
instance, computer simulations produce a lot of fruitful
results such as in Lattice
QCD\cite{QCD}, but, in addition to the consumption of money
(as well as time), there still
lacks something to put symmetry onto the lattice; chiral
symmetry is a
well-known example\cite{CHIRAL}. An advantage in path
integration, contrary to the operator
formalism, is that we can easily switch from one variable to
other by means
of some change of variables, which would open a new possibility.
The auxiliary field is
considered as one of these variables, and was introduced into
the model by Gross and
Neveu\cite{GN}.

The Gross-Neveu model is a two-dimensional
four-fermion model inspired by the work by Nambu and
Jona-Lasinio\cite{NJL};
\begin{equation}
 {\cal L} =   \overline{\psi} i {\partial \hspace{-1.2ex}/}
 \psi + { g^2 \over
2} \left( \overline{\psi} \psi \right)^2   \  ,
\label{GN}
\end{equation}
where $\psi$ has an $N$-component. After some discussions,
they proposed an equivalent
Lagrangian,
\begin{equation}
{\cal L'} =   \overline{\psi} i {\partial \hspace{-1.2ex}/}
\psi - { \sigma^2 \over 2}
- g  \overline{\psi} \psi \sigma  \  .
\end{equation}
Here $\sigma$ has no kinetic term and is eliminated by using
the equation of
motion yielding to the original Lagrangian eq.(\ref{GN}).
In this sense, we call
$\sigma$ as an auxiliary field.

The scenario is more easily understood by means of path
integral\cite{RIVERS}: the
partition function (in an imaginary temperature) reads
\begin{equation}
 Z\equiv \int {\cal D}\psi {\cal
D}\overline{\psi} \exp\left[ i \int d^2 x
\left( \overline{\psi} i {\partial
\hspace{-1.2ex}/}
\psi + { g^2 \over
2} \left( \overline{\psi} \psi \right)^2 \right) \right]  \ .
\label{generating}
\end{equation}
Here introducing the auxiliary field $\sigma$ in terms of
the Gaussian integrals,
such that
\begin{equation}
1= \int {\cal D} \sigma \exp \left[  - i \int d^2 x
{ 1 \over 2}  \left( \sigma +  g
\overline{\psi}
\psi \right)^2 \right]     \ ,
\end{equation}
and inserting into eq.(\ref{generating}) we find
\begin{equation}
 Z =\int \! d \sigma {\cal D}\psi {\cal
D}\overline{\psi} \  e^{ i \int d^2 x  {\cal L'}}=
\int \! d \sigma {\cal D}\psi
{\cal D}\overline{\psi} \exp \! \left[ i \int d^2 x
\left( \overline{\psi} i {\partial
\hspace{-1.2ex}/}  \psi - { \sigma^2 \over 2}  - g
\overline{\psi} \psi \sigma \right)
\right]
\ .  \label{GNaux}
\end{equation}
Similar techniques are utilized everywhere nowadays also
for a boson quartic
interaction\cite{GN2} instead of the four-fermi interaction.
The nomenclature for
$\sigma$-field is, therefore, various;
the mean-field\cite{ANNA}, the
Hubbard-Stratonovich field\cite{FRAD} in solid state physics.

In the actual case, we treat the partition function
(which can be obtained through $t
\mapsto it$):
\begin{equation}
 Z(T) \equiv  \int \! d \sigma {\cal D}\psi
{\cal D}\overline{\psi} \exp \! \left[  -
\int_0^T dt \int dx \left( \overline{\psi}
{\partial
\hspace{-1.2ex}/}  \psi + { \sigma^2 \over 2} +  g
\overline{\psi} \psi \sigma \right)
\right]
\ ,  \label{}
\end{equation}
where the anti-periodic boundary condition for the fermi
field, $\psi(T,
x)=-\psi(0, x)$, should be understood. We then integrate
out the fermion
field to find
\begin{equation}
\begin{array}{c}
\displaystyle{ \hspace{0ex}  Z(T) =  \int \! d \sigma
\exp \! \left[ - \int d^2 x   {\sigma^2\over 2}
   + N \ln \det \left( {\partial \hspace{-1.2ex}/}+ g
   \sigma \right)
\right] \equiv  \int \! d \sigma e^{- S[\sigma] }
\  ,    }                  \\
\noalign{\vspace{1ex} }
\displaystyle{ \hspace{0ex} S[\sigma] \equiv \int d^2 x
{\sigma^2\over 2}
   - N \ln \det \left( {\partial \hspace{-1.2ex}/}+ g
   \sigma \right) \ . }
\end{array}
\end{equation}
Since we look for a vacuum with $T\mapsto \infty$,
we should find
a constant solution $\sigma_0$ in the equation of motion,
\begin{equation}
  \left. {\delta S \over \delta \sigma(x)}\right|_{\sigma_0}
  = 0      \ ,
\end{equation}
which gives the gap equation;
\begin{equation}
 \sigma_0 =  2Ng^2   \int {d^2 k \over (2 \pi)^2} \
 { \sigma_0  \over k^2 +
(g \sigma_0)^2 }  \ .
\end{equation}
If $\sigma_0$ is non-zero then the dynamically symmetry
breaking occurs: this is the
end of the usual story. Indeed, the recipe is legitimated
if the number of fermion
species becomes infinite; $N \rightarrow \infty$. However,
it is not the case for most of
actual situations: $N$ is finite or even $1$.
We wish to know ``how accurate is it when $N=1$?'', which is
one of the motivation of this work.

Moreover there is an alternative motivation:
performing the WKB approximation in
the Double-Well potential we must follow instanton
calculations\cite{SC}, which
however is very cumbersome as well as tedious.
The simpler is the better in any approximation:
once
introduce an auxiliary field we could avoid such
troublesome. To
clarify these issues, we pick up the case of quartic
coupling of bosonic field for simplicity.

The paper is organized as follows: in
\S II a simple model calculation is performed for
the integral expression. Here we realize the
importance
of the loop-expansion with respect to the auxiliary
field(variable) and can find a more accurate result is
obtained if taking a higher loop correction into account
when the coupling, $g$, is
small. However, when $g$ goes larger, higher loops cannot
always improve a situation.
We then proceed to the quantum mechanical model in
\S III, where we compare our results with those obtained
numerically to find that the
2-loop correction gives a $4\%$-error for
$10^{-3}< g^2 < 10^3$ except $g^2 \sim
O\! \left(10^{-1}\right)$ in the Double-Well case.
The final section is devoted to a discussion.

\section{Simple (0-dimensional) Model}
The starting point is
\begin{equation}
  I  \equiv   \int_{-\infty}^{\infty}
  { dx \over \sqrt{2 \pi} } \exp \left[ - {\omega^2
\over 2} x^2 - { g^2 \over 8} x^4 \right]   \ .
\label{simple}
\end{equation}
The integral is expressed as
\begin{equation}
I   =  \sqrt{ \pi | \omega^2 |  \over 4 g^2 }
e^{ \omega^4 / 4
g^2}
\left\{ {\rm I}_{-1/4} \! \left( \!
{\omega^4 \over 4g^2} \right) - \epsilon(\omega^2)
{\rm I}_{1/4} \!
\left( \! {\omega^4 \over 4g^2} \right)  \right\}  \ ,
\label{integral}
\end{equation}
where
$$
   \epsilon(x) =  \left\{  \begin{array}{c}  +1
   \  :   \  x  >  0  \\
      -1  \  :    \ x  <    0     \end{array}
  \right.   \  ,
$$
and ${\rm I}_{\alpha} (x) $ is the modified Bessel function.
There are two
cases depending on a sign $\omega^2$:

\vspace{2ex}

\noindent CASE(i); $\omega^2 > 0$;
(0-dimensional Anharmonic-Oscillator),
\begin{equation}
\begin{array}{l}
\displaystyle{ \hspace{0ex} I =
\sqrt{ \omega^2  \over 2 \pi g^2 }  e^{ \omega^4 / 4 g^2}
\  {\rm
K}_{1/4} \!
\left( \! {\omega^4 \over 4g^2} \right)     }                  \\
\noalign{\vspace{1ex} }
\displaystyle{ \hspace{1ex}
\stackrel{g^2 \rightarrow 0}{ \sim} 1 + O(g^2)  \  ; }
\end{array}
\end{equation}
where ${\rm K}_{\alpha} (x)$ also is the modified Bessel function. And

\vspace{2ex}

\noindent CASE(ii); $\omega^2 < 0 $; (0-dimensional Double-Well),
\begin{equation}
     I \stackrel{g^2 \rightarrow 0}{ \sim} \sqrt{2}  \ e^{ \omega^4
/ 2 g^2}
\  .                        
\end{equation}
Here it should be noted that $g^2=0$ is the essential singularity,
that is to say, the
Double-Well case is non-Borel summable.

Now introduce an auxiliary field such that
\begin{equation}
1= \int_{-\infty}^\infty { dy \over \sqrt{2 \pi} }
\exp\left[ - {1 \over 2} \left( y
+ i g { x^2 \over 2} \right)^2 \right]  \
\end{equation}
so as to erase the $x^4$ term when inserted into eq.(\ref{simple}),
yielding
\begin{equation}
\begin{array}{l}
\displaystyle{ \hspace{0ex} I =  \int_{-\infty}^\infty
{ dx  \over \sqrt{2 \pi} } { dy \over
\sqrt{2 \pi} } \exp
\left[  -
{1
\over 2} (\omega^2 + igy)x^2    - {y^2 \over 2} \right]  }
\\
\noalign{\vspace{1ex} }
\displaystyle{ \hspace{2ex}= \int_{-\infty}^\infty
{ dy \over \sqrt{2 \pi} } \  (\omega^2
+ig y)^{-1/2} \ e^{ - {y^2
\over 2 }} =  \int_{-\infty}^\infty { dy \over \sqrt{2 \pi} }
\exp\left[ - {y^2
\over 2 } -  {1
\over 2 } \ln \left( \omega^2 +ig
y \right) \right]  \  .   }
\end{array}                           \label{xintegral}
\end{equation}
We rewrite the final expression as
\begin{equation}
\begin{array}{c}
\displaystyle{I  =  \int_{-\infty}^\infty { dy \over \sqrt{2 \pi} }
 \left. \exp\left[ - { S(y) \over a } \right]
 \right|_{a=1}   \ ,   }       \\
\noalign{\vspace{1ex} }
\displaystyle{ \hspace{0ex} S(y) \equiv {1 \over 2} \ln
(\omega^2 + igy) + {y^2 \over 2 }   \  , }
\end{array}
\label{final}
\end{equation}
where we have introduced a parameter, $a$,
which must be put unity in
the final stage. We call $a$ as the loop-expansion
parameter. Next, assume the solution
of $S'(y)=0$ as $y_{ {}_{0} }$;
\begin{equation}
 S'(y) = y +  { ig \over 2( \omega^2 + igy) } = 0    \  ,
\end{equation}
which can be expressed as
\begin{equation}
\displaystyle{ {\rm \Omega}^2
-\omega^2 = { g^2 \over 2
{\rm \Omega}^2} \ ;  }    \label{equationa}
\end{equation}
where ${\rm \Omega}^2$ should obey
\begin{equation}
\displaystyle{ \hspace{0ex}   {\rm \Omega}^2
\equiv
\omega^2 + igy_{ {}_{0} } > 0   \  ;  }
\label{equationb}
\end{equation}
since the Gaussian integral of $x$ eq.(\ref{xintegral})
must exist. Then perform
the saddle point method around
$y_{ {}_{0} }$ to give
\begin{equation}
\begin{array}{l}
\displaystyle{I= \int_{-\infty}^\infty { dy \over \sqrt{2 \pi} }
\ \exp \! \left[ - { S(y_{ {}_{0} })
 \over  a}  - { S^{(2)}(y_{ {}_{0} })
 \over 2  a} (y-y_{ {}_{0} })^2  - { S^{(3)}(y_{ {}_{0} })
 \over 3!  a} (y-y_{ {}_{0} })^3   \right. }    \\
\noalign{\vspace{1ex} }
\displaystyle{ \hspace{18ex} \left. \left. - { S^{(4)}(y_{ {}_{0} })
 \over 4!  a}  (y-y_{ {}_{0} })^4 - \cdots  \right]
 \right|_{ a =1}  \  . }
\end{array}
\end{equation}
Making a change of variable, $y-y_{ {}_{0} } \mapsto y/ \sqrt{a}$,
we obtain
\begin{equation}
\begin{array}{l}
\displaystyle{I= e^{-S_0/ a } \left.
\int_{-\infty}^\infty
\sqrt{  a  \over
 2 \pi    }  \   dy
\exp \! \! \left[  -  { S^{(2)}_0 \over 2 } y^2  -
\sqrt{ a } \  { S^{(3)}_0 \over 3! } y^3  - a
{ S^{(4)}_0 \over 4!  } y^4 - \cdots \right]
\right|_{ a =1}  }
\\
\noalign{\vspace{1ex} }
\displaystyle{  \hspace{2ex}  \cong   \left.
e^{ - S_0/ a }
 \sqrt{ a
\over  2 \pi   }  \int_{-\infty}^\infty  dy
\   e^{  -   S^{(2)}_0
 y^2 /2  } \left( 1  -  a
\left\{  { S^{(4)}_0 \over 4!  } \  y^4  -   { {S^{(3)}_0}^2
\over 2 (3!)^2 } \  y^6  \right\}  + O(a^2) \right)
\right|_{a=1}  \ , }
\end{array}
\end{equation}
where we have written
\begin{equation}
S^{(n)} \! \left(y_{ {}_{0} }\right) \equiv S^{(n)}_0 \ .
\end{equation}
$a^{L-1}$ term is called the $L$-loop term ($L=0$ is the tree term.)
From eq.(\ref{equationa}),
\begin{equation}
 {\rm \Omega}^2 = { \omega^2 + \sqrt{\omega^4 + 2g^2} \over 2}
 \ ,  \label{0solution}
\end{equation}
\begin{equation}
\begin{array}{ll}
\displaystyle{S_0 =  - { \left( {\rm \Omega}^2 - \omega^2 \right)^2
\over 2 g^2 } = - {
g^2 \over 8 {\rm \Omega}^4}
\ ,  } &
\displaystyle{S_0^{(2)} =
1 + { g^2 \over 2 {\rm \Omega}^2 }  \ ,   }  \\
\noalign{\vspace{1ex} }
\displaystyle{S_0^{(3)} =
-i  { g^3 \over  {\rm \Omega}^6 } \ , \quad S_0^{(4)} = -3  { g^4
\over  {\rm \Omega}^8 } \  ,  }  &
\displaystyle{   S_0^{(6)} =   { 5! g^6
\over  2 {\rm \Omega}^{12} }  \ .    }
\end{array}
\end{equation}
Using these and performing elementary integrals, we obtain
\begin{equation}
\begin{array}{l}
\displaystyle{I =  \exp \left( { g^2 \over  8
{\rm \Omega}^4} \right)   \sqrt{ {\rm \Omega}^2 \over
{\rm \Omega}^4 + { g^2\over 2}
} \left( 1+   { 3g^4   \over 8  \left(
{\rm \Omega}^4 + { g^2\over 2} \right)^2 }  \right. }       \\
\noalign{\vspace{1ex} }
\displaystyle{ \hspace{6ex}   \left.  -
{ 35g^6   \over 24  \left(  {\rm \Omega}^4
+ { g^2\over 2} \right)^3 }  +
{ 329 g^8   \over 128  \left(  {\rm \Omega}^4
+ { g^2\over 2} \right)^4  }   -
{ 105 g^{10}   \over 64  \left(  {\rm \Omega}^4
+ { g^2\over 2} \right)^5 }  +   O(\mbox{4-loop})   \right)  \  . }
\end{array}
\end{equation}
Stated as above, we assign
\begin{equation}
\begin{array}{l}
\displaystyle{ \   I_{\rm tree}\equiv  \
\exp \left( { g^2 \over  8
{\rm \Omega}^4} \right) \ ,   }       \\
\noalign{\vspace{1ex} }
\displaystyle{ \hspace{0ex}  I_{\rm 1-loop}\equiv
I_{\rm tree}  \sqrt{ {\rm \Omega}^2 \over {\rm
\Omega}^4 + { g^2\over 2}  }  \ ,  }  \\
\noalign{\vspace{1ex} }
\displaystyle{ \hspace{0ex}  I_{\rm 2-loop}
\equiv  I_{\rm 1-loop} \left( 1+   { 3g^4   \over 8  \left(
{\rm \Omega}^4 + { g^2\over 2} \right)^2 }  -
{ 5g^6   \over 24  \left(  {\rm \Omega}^4 +
{ g^2\over 2} \right)^3 }   \right) \ ,    }  \\
\noalign{\vspace{1ex} }
\displaystyle{ \hspace{0ex}  I_{\rm 3-loop} \equiv
\exp \left( { g^2 \over  8
{\rm \Omega}^4} \right)
\sqrt{ {\rm \Omega}^2 \over {\rm \Omega}^4 + { g^2\over 2}
} \left( 1+   { 3g^4   \over 8  \left(
{\rm \Omega}^4 + { g^2\over 2} \right)^2 }  \right.    }  \\
\noalign{\vspace{1ex} }
\displaystyle{ \hspace{8ex}
\left. -   { 35g^6   \over 24  \left(  {\rm \Omega}^4
+ { g^2\over 2} \right)^3 }  +
{ 329 g^8   \over 128  \left(  {\rm \Omega}^4
+ { g^2\over 2} \right)^4  }   -
{ 105 g^{10}   \over 64  \left(  {\rm \Omega}^4
+ { g^2\over 2} \right)^5 }    \right) \ .   }
\end{array}
\label{expression}
\end{equation}

\begin{figure}
\centering\leavevmode
\epsfysize=6cm
 \epsfbox{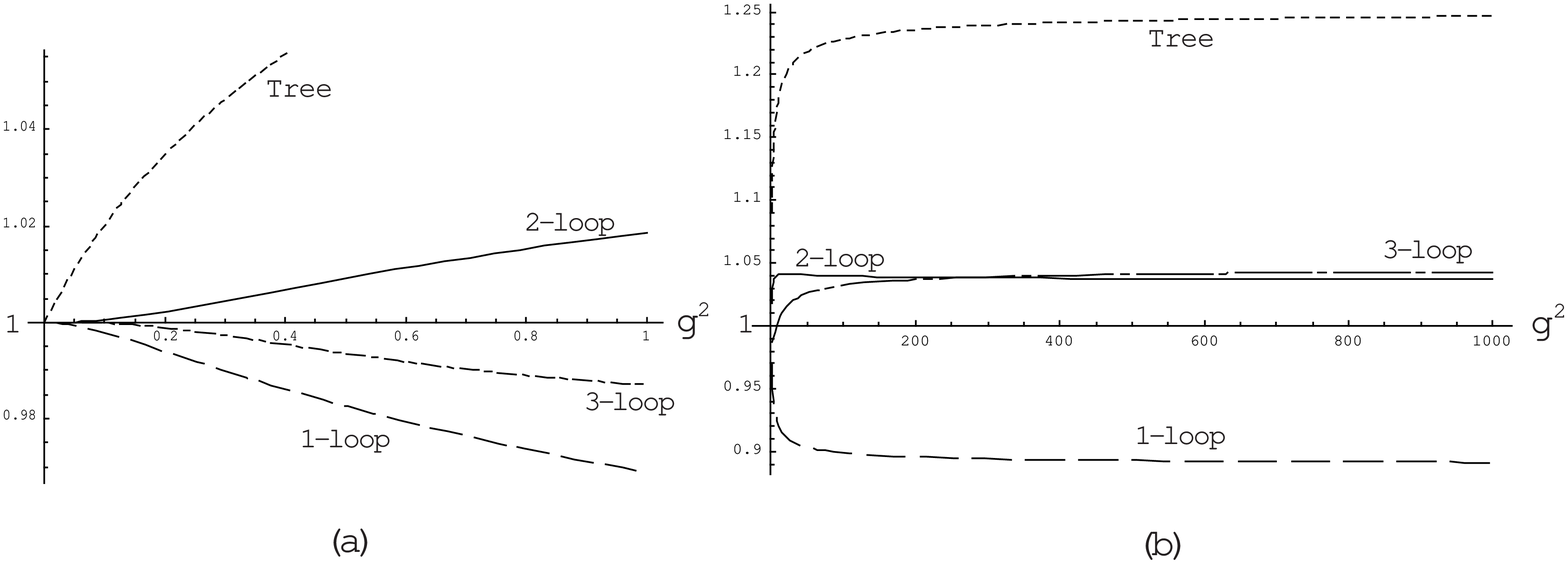}
\caption{0-dimensional Anharmonic Case:
(a); $g^2 \leq 1$. (b);  $g^2
>1$.  Dotted line: $I_{\rm Tree}/\mbox{Exact}$.
Dashed line: $I_{\rm
1-loop}/\mbox{Exact}$.
Solid line: $I_{\rm
2-loop}/\mbox{Exact}$.
Dash-dotted line: $I_{\rm 3-loop}/\mbox{Exact}$.}
\label{FirstFigure}
\end{figure}

Let us analyze the individual case:
\begin{itemize}
\item CASE(i); 0-dimensional Anharmonic-Oscillator.
Put  $\omega^2 \mapsto 1$ so that
eq.(\ref{0solution}) reads
\begin{equation}
{\rm \Omega}^2 = { \sqrt{1+ 2g^2} + 1\over 2}   \  . 
\end{equation}
We plot the ratio of $I_{\rm L-loop}$(L=0,1,2,3)
to the exact value in Fig.(\ref{FirstFigure}).
(a) shows the case of $g^2 \leq 1$ and (b) of $g^2 >1$.
Details are seen in table (a).
\end{itemize}

\vspace{0ex}

\begin{center}
{\small
\begin{tabular}{| c | c | c | c | c | c | }  \hline
$\displaystyle{ g^2 }$  &  Exact  &
$\begin{array}{c}
\mbox{Tree} \\
   \mbox{Tree/ Ex.}
\end{array}$ & $\begin{array}{c}
\mbox{1-loop} \\
  \mbox{1-loop/ Ex.}
\end{array}$  &
$\begin{array}{c}
\mbox{2-loop} \\
  \mbox{2-loop / Ex.}
\end{array}$  &
$\begin{array}{c}
\mbox{3-loop} \\
  \mbox{3-loop / Ex.}
\end{array}$  \\   \hline
$10^{-3}$ & 0.9996 & $\begin{array}{c}
0.9999 \\   1.00
\end{array}$ & $\begin{array}{c}
0.9996 \\   1.
\end{array}$  & $\begin{array}{c}
0.9996 \\   1.
\end{array}$  &  $\begin{array}{c}
0.9996 \\   1.
\end{array}$   \\  \hline
$10^{-2}$  & 0.9963 & $\begin{array}{c}
0.9988  \\   1.00
\end{array}$  & $\begin{array}{c}
 0.9963  \\   1.
\end{array}$  &  $\begin{array}{c}
 0.9963  \\   1.
\end{array}$   &  $\begin{array}{c}
 0.9963  \\   1.
\end{array}$  \\   \hline
$10^{-1}$  & 0.9685 & $\begin{array}{c}
0.9881  \\   1.02
\end{array}$   &  $\begin{array}{c}
0.9664  \\   1.00
\end{array}$ & $\begin{array}{c}
0.9690  \\   1.00
\end{array}$ & $\begin{array}{c}
0.9683  \\   1.00
\end{array}$   \\  \hline
$1 $  & 0.8386 & $\begin{array}{c}
  0.9149  \\   1.1
\end{array}$ &  $\begin{array}{c}
 0.8125 \\   0.97
\end{array}$  &  $\begin{array}{c}
 0.8541 \\   1.02
\end{array}$  &   $\begin{array}{c}
 0.8277  \\   0.99
\end{array}$ \\  \hline
$10 $  & 0.5954  & $\begin{array}{c}
 0.7027    \\   1.18
\end{array}$ &  $\begin{array}{c}
0.5484  \\   0.92
\end{array}$  &  $\begin{array}{c}
 0.6195\\  1.04
\end{array}$   &  $\begin{array}{c}
 0.5976\\  1.00
\end{array}$   \\
\hline
$10^{2}$  &   0.3672 & $\begin{array}{c}
  0.4510   \\   1.23
\end{array}$ &  $\begin{array}{c}
  0.3300 \\   0.90
\end{array}$  &  $\begin{array}{c}
 0.3817 \\   1.04
\end{array}$   &  $\begin{array}{c}
 0.3790  \\   1.03
\end{array}$   \\
\hline
$10^{3}$  &  0.2131 & $\begin{array}{c}
   0.2656  \\   1.25
\end{array}$ &  $\begin{array}{c}
 0.1899 \\   0.89
\end{array}$  &  $\begin{array}{c}
0.2210  \\   1.037
\end{array}$  &  $\begin{array}{c}
0.2222  \\   1.043
\end{array}$  \\
\hline
 \end{tabular}
}
\vspace{2ex}

table(a)

\end{center}

\begin{figure}
\centering\leavevmode
 \epsfysize=6cm
\epsfbox{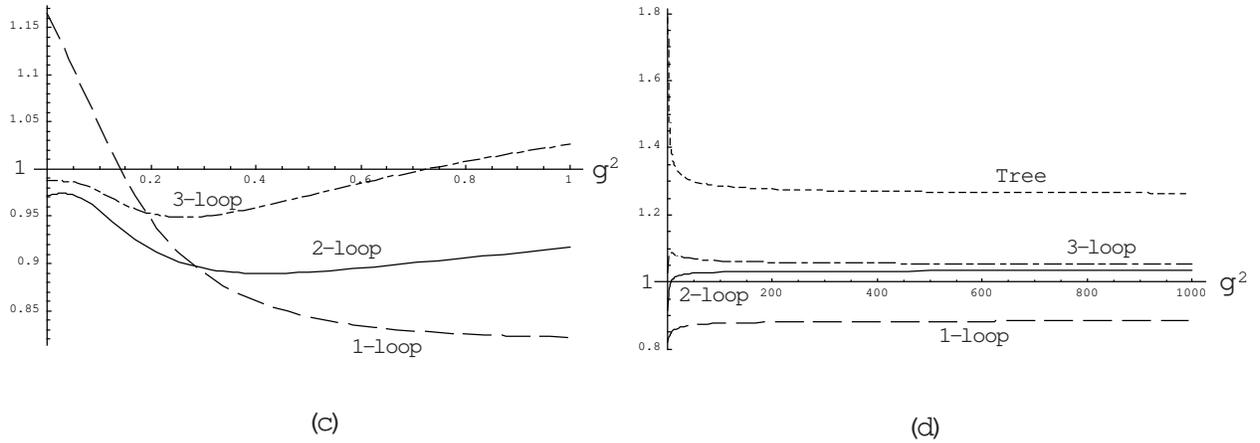}
\caption{0-dimensional Double-Well Case:
(c); $ g^2 \leq 1$. (d);
$ g^2 >1$.
Dotted line: $I_{\rm Tree}/\mbox{Exact}$:
this is omitted in (c);
because of a
large deviation. Dashed line:
$I_{\rm 1-loop}/\mbox{Exact}$. Solid line: $I_{\rm
2-loop}/\mbox{Exact}$. Dash-dotted line:
$I_{\rm 3-loop}/\mbox{Exact}$.}
\label{SecondFigure}
\end{figure}

\vspace{2ex}
\begin{itemize}
\item CASE (ii); 0-dimensional Double-Well.  Put $\omega^2
\mapsto -1$  so that
eq.(\ref{0solution}) reads
\begin{equation}
{\rm \Omega}^2 = { \sqrt{1+ 2g^2} -1  \over 2}   \  .   
\end{equation}
We plot the same ratio as the above in Fig.(\ref{SecondFigure}).
However, in (c) we have omitted the tree graph because of the large
deviation. Details are shown in table (b).
\end{itemize}

\vspace{0ex}

\begin{center}
{\small
\begin{tabular}{| c | c | c | c | c | c | }  \hline
$\displaystyle{ g^2 }$ & Exact & $ \hspace{-1ex} \begin{array}{c}
\mbox{Tree}  \\
\mbox{Tree/Ex.}
\end{array}  \hspace{-1ex} $  &  $ \hspace{-1ex} \begin{array}{c}
\mbox{1-loop}  \\
 \mbox{1-loop/Ex.}
\end{array} \hspace{-1ex} $ & $ \hspace{-1ex} \begin{array}{c}
\mbox{2-loop}  \\
 \mbox{2-loop/Ex.}
\end{array}  \hspace{-1ex}  $   & $ \hspace{-1ex} \begin{array}{c}
\mbox{3-loop}  \\
 \mbox{3-loop/Ex.}
\end{array}  $  \\  \hline
$10^{-3}$ & $ 1.986\! \times \! 10^{217} $ &
$  \begin{array}{c} 1.035\! \times\! 10^{219}
 \\     52.1
\end{array}   $ & $  \begin{array}{c}
2.313\!\times\! 10^{217}
\\   1.17
\end{array}   $  & $    \begin{array}{c}
1.930\! \times\! 10^{217}
\\   0.97
\end{array}   $  & $   \begin{array}{c}
1.961 \! \times\! 10^{217}
\\   0.98
\end{array}   $    \\  \hline
$10^{-2}$ &  $ 7.360 \! \times\! 10^{21}  $ &
$  \begin{array}{c} 1.210\! \times \! 10^{23}
 \\   16.44
\end{array} $ &  $ \begin{array}{c}
8.495  \! \times \! 10^{21}
\\    1.15
\end{array} $ &  $  \begin{array}{c}
 7.162 \! \times \! 10^{21}
\\    0.97
\end{array}$   &  $  \begin{array}{c}
 7.270 \! \times \! 10^{21}
\\    0.98
\end{array}$ \\   \hline
$10^{-1}$  & $   2.208 \! \times \! 10^{2} $ &
$  \begin{array}{c} 1.107 \! \times \! 10^{3}  \\   5.01
\end{array} $   &  $  \begin{array}{c}
 2.311 \! \times \! 10^{2}
\\   1.05
\end{array}  $  &  $ \begin{array}{c}
 2.112\! \times \! 10^{2}
  \\    0.96
\end{array}$  &  $ \begin{array}{c}
 2.156 \! \times \! 10^{2}
  \\    0.98
\end{array}$   \\  \hline
$1 $  & 2.350 & $\begin{array}{c}
4.202 \\  1.79
\end{array}$ &  $\begin{array}{c}
1.932 \\  0.82
\end{array}$  &  $\begin{array}{c}
2.155 \\   0.92
\end{array}$  &  $\begin{array}{c}
2.411  \\     1.03
\end{array}$    \\
\hline
$10 $  & 0.8074  & $\begin{array}{c}
1.103   \\   1.37
\end{array}$ &  $\begin{array}{c}
0.6897  \\  0.85
\end{array}$  &  $\begin{array}{c}
0.8137\\   1.01
\end{array}$    &  $\begin{array}{c}
0.8768 \\   1.09
\end{array}$    \\
\hline
$10^{2}$  &   0.4040 & $\begin{array}{c}
 0.5196   \\   1.29
\end{array}$ &  $\begin{array}{c}
 0.3542 \\  0.88
\end{array}$  &  $\begin{array}{c}
0.4159 \\  1.03
\end{array}$  &  $\begin{array}{c}
0.4295  \\  1.06
\end{array}$    \\
\hline
$10^{3}$  &  0.2196 & $\begin{array}{c}
0.2777  \\  1.26
\end{array}$ &  $\begin{array}{c}
0.1942 \\  0.88
\end{array}$  &  $\begin{array}{c}
0.2270  \\   1.03
\end{array}$  &  $\begin{array}{c}
0.2312  \\   1.05
\end{array}$    \\
\hline
\end{tabular}

}
\vspace{2ex}

table(b)

\end{center}
\vspace{2ex}

From the figures, (\ref{FirstFigure}) and (\ref{SecondFigure}),
it should be noted that the higher loop corrections improve the result
all the time when $g^2 \leq 1$ but not in the strong coupling
region as is seen from
the graphs (b) and (d). Details for the values are listed in
the tables (a) and (b). This fact implies
that the loop-expansion is merely an asymptotic expansion.
In the Anharmonic Oscillator case,
especially the result is satisfactory: the 2-loop result gives
a 4\%-error for $10^{-3}< g^2 <
10^3$.  In the Double-Well case, the 3-loop spoils the result
at $1< g^2$ but gives a better result at $g^2
<1$.  However, it is still remarkable that the error, under
the 2-loop, remains within $\sim 8\%$ for a
huge coupling region, $10^{-3}< g^2 < 10^3$.

The essential role of the loop-expansion should finally be remarked:
if we stop at the $g^4$ term in
the 2- or 3-loop expression eq.(\ref{expression}), the result deviates
far away from the true value.
Therefore we must abandon the coupling constant expansion in the
auxiliary field method.


\section{The Quantum Mechanical Model}
Encouraged by the foregoing results, in this section we analyze
the quantum mechanical
model:
\begin{equation}
H = { p^2 \over 2} + {\omega^2 \over 2} x^2 + { g^2 \over 8}
x^4  \  .                            
\end{equation}
Here again cases are classified into (i) $\omega^2 > 0$;
Anharmonic-Oscillator. And (ii) $\omega^2<0$; Double-Well.
The partition function is given
by
\begin{equation}
\begin{array}{l}
\displaystyle{ \hspace{0ex} Z(T) =   {\rm Tr} \ e^{-TH}     }
\\
\noalign{\vspace{1ex} }
\displaystyle{ \hspace{6ex} = \int {\cal D} x
\exp\left.\left[ - \int_0^T \! \! dt \left(
 { {\dot{x}}^2 \over 2}  + {\omega^2
\over 2} x^2  + { g^2
\over 8} x^4  \right) \right] \right|_{x(T)=x(0)}
\ , }
\end{array}
\end{equation}
where $x(T)=x(0)$ designates the periodic boundary condition.
Here and hereafter we put $\hbar
\mapsto 1$ and use the continuous representation,
\begin{equation}
 {\cal D} x \equiv \lim_{N \rightarrow \infty}
 \prod_{j=1}^{N}  { dx_j
\over \sqrt{ 2 \pi \Delta t}  } \ ;   \qquad
\Delta t \equiv { T \over N } \
.
\end{equation}
Introducing an auxiliary field in terms of the Gaussian identity,
\begin{equation}
  1=  \int {\cal D} y
\exp \! \left[ - \int_0^T \! \! dt
 \  {1 \over 2} \left( y + {igx^2 \over 2} \right)^2 \right]  \ ,
\end{equation}
so as to erase the quartic term, we obtain
\begin{equation}
Z(T) =   \int \!  {\cal D} x {\cal D} y \exp  \! \left[   - \!
\int_0^T
\!
\! dt \!
 \left(  {{\dot{x}}^2 \over 2} + \left(\! \omega^2 \!
 + igy\right) { x^2 \over 2}  +{y^2
\over 2} \right)  \! \right]    \ ,
\label{x-int}
\end{equation}
which becomes, after the integration with respect to $x$, to
\begin{equation}
\begin{array}{c}
\displaystyle{ \hspace{0ex} Z(T) = \int \!  \! {\cal D} y
\exp \! \left[ \!   -  \int_0^T \! \! dt \  {y^2 \over 2}
+ {1 \over 2}  \ln \det \! \left( \! - {d^2 \over dt^2 } +
\! \omega^2 \!  +
igy \!  \right)   \right]  }             \\
\noalign{\vspace{1ex} }
\displaystyle{ \hspace{0ex} \equiv \int \!  \! {\cal D} y
\left. \exp \! \left( - { S[y]
\over a }
\right)  \right|_{a=1}  \ ,  }
\end{array}
\end{equation}
where
\begin{equation}
 S[y] \equiv   \int_0^T \! \! dt \  {y^2 \over 2} +
 {1 \over 2}  \ln \det \! \left( \! - {d^2 \over dt^2 } + \!
\omega^2 \!  + igy \!  \right)      \ ,
\end{equation}
and again the loop-expansion parameter,$a$, has been introduced.

Write a solution of
the equation of motion, $S'[y] = \delta S[y]  / \delta y(t) = 0$,
${y}_{ {}_{0} }(t)$, giving the gap
equation;
\begin{equation}
{y}_{ {}_{0} }(t) + { ig \over 2} G(t,t)  = 0 \  ,
\label{}
\end{equation}
which can be rewritten as
\begin{equation}
 {{\rm \Omega}(t)}^2 - \omega^2 =  { g^2
\over 2} G(t, t)     \ ,
\label{gap}
\end{equation}
where
\begin{equation}
 {\rm \Omega}(t)^2 \equiv
\omega^2 + ig {y}_{ {}_{0} }(t)  \ .
\end{equation}
Here the Green's function,$G(t, t')$, obeys
\begin{equation}
\left( - { d^2 \over dt^2} +{\rm  \Omega}(t)^2
\right)  G(t, t') = \delta(t-t')        \ .
\label{green}
\end{equation}
Again it should be noted that
\begin{equation}
 {\rm  \Omega}(t)^2 > 0 \ ,
\end{equation}
due to the existence of the Gaussian integration of $x$
in eq.(\ref{x-int}).

\begin{figure}
\centering\leavevmode
\epsfysize=8cm  \epsfbox{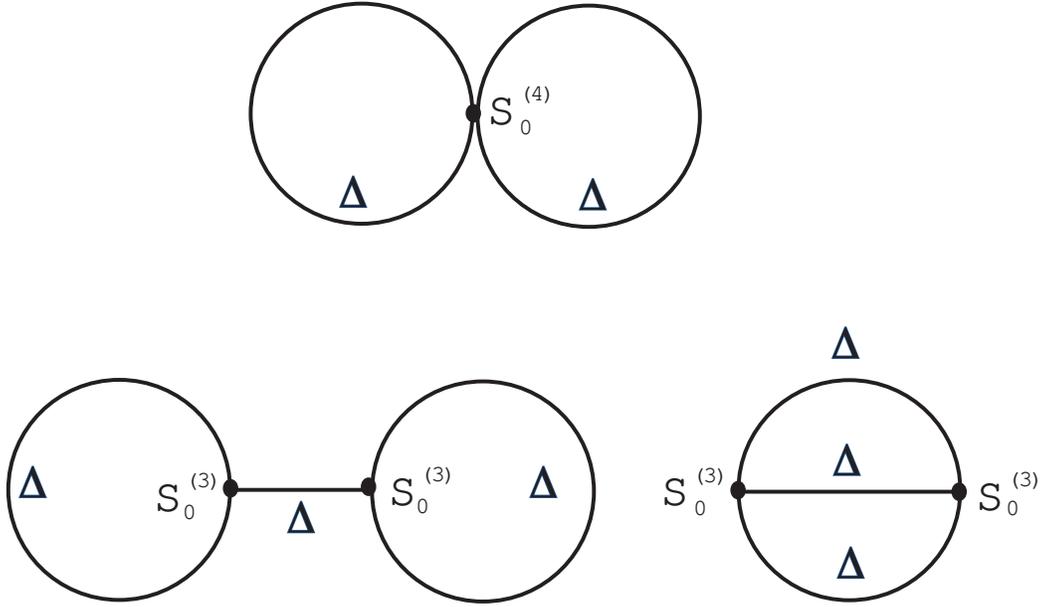}
\caption{Formal 2-loop Graphs:
$\Delta$ denotes the
propagator of the auxiliary field.}
\label{ThirdFigure}
\end{figure}

Now
$S[y]$ is expanded around ${y}_{ {}_{0} }$ (or ${\rm
\Omega}(t)^2$) such that
\begin{equation}
S[y]=  S_0 + { 1 \over 2} (y- {y}_{ {}_{0}})^2 \cdot S^{(2)}_0 +
{ 1 \over 3! }  (y- {y}_{
{}_{0}})^3 \cdot  S^{(3)}_0  + \cdots   \  ,
\end{equation}
where abbreviations
\begin{equation}
 S^{(n)}_0 \equiv \left. { \delta^n S \over \delta y(t_1)
\delta y(t_2) \cdots \delta y(t_n)   }\right|_{y= {y}_{ {}_{0}} } \ ,
\end{equation}
and
\begin{equation}
\begin{array}{l}
\displaystyle{ \hspace{0ex} ( y- {y}_{ {}_{0}} )^n \cdot S^{(n)}_0  }
\\
\noalign{\vspace{1ex} }
\displaystyle{ \hspace{0ex}  \equiv \int  dt_1 dt_2 \cdots dt_n
(y- {y}_{ {}_{0}})(t_1)    (y- {y}_{{}_{0}})(t_2)
\cdots  (y- {y}_{ {}_{0}})(t_n)
\left. { \delta^n S \over \delta y(t_1)
\delta y(t_2) \cdots \delta y(t_n)   }\right|_{y= {y}_{ {}_{0}} }
\ ,  }
\end{array}
\end{equation}
have been adopted. Shifting and scaling the integration variables
as before, we obtain
\begin{equation}
 Z(T) = e^{ -  S_0 / a } \left.
\int  \! { \cal D} y  \exp  \!  \left[  - { 1 \over 2}  \Delta^{-1}
\cdot y^2 - \sqrt{a} \
{ y^3
\over 3! }  \cdot S^{(3)}_0 - a \ { y^4 \over 4! }
\cdot  S^{(4)}_0 - \cdots \right]
\right|_{a=1}
\ ,
\end{equation}
where we have written $S^{(2)}_0 \mapsto \Delta^{-1}$ which reads
explicitly
\begin{equation}
\Delta^{-1}(t_1, t_2) \equiv  \left.{ \delta^2
S \over \delta y(t_1)  \delta y(t_2)  }\right|_{y= {y}_{ {}_{0}} }
= \delta(t_1-t_2) + { g^2
\over 2} G(t_1,t_2) G(t_2, t_1)   \ .
\end{equation}
Moreover,
\begin{equation}
\begin{array}{l}
\displaystyle{ \hspace{0ex}  S^{(3)}_0  = - { i g^3
\over 2} \Bigg\{  G(t_1, t_2) G(t_2, t_3) G(t_3, t_1)
+  G(t_1, t_3) G(t_3, t_2) G(t_2,
t_1)   \Bigg\}   \ , }                  \\
\noalign{\vspace{1ex} }
\displaystyle{ \hspace{0ex} S^{(4)}_0 = -  g^4
 \Bigg\{  G(t_1, t_2) G(t_2, t_3) G(t_3, t_4) G(t_4, t_1) +
 G(t_1, t_2) G(t_2, t_4) G(t_4,
t_3) G(t_3, t_1)    } \\
\noalign{\vspace{1ex} }
\displaystyle{ \hspace{25ex}
+ G(t_1, t_3) G(t_3, t_2) G(t_2, t_4) G(t_4, t_1) \Bigg\} \ .
}
\end{array}
\end{equation}
From these we have
\begin{equation}
\begin{array}{l}
\displaystyle{ \hspace{0ex}   { Z(T) }_{ {\rm Tree} }
\equiv  \exp\left( - S_0 \right)  \ ,
}
\\
\noalign{\vspace{1ex} }
\displaystyle{ \hspace{0ex}   {Z(T)}_{ {\rm 1-loop } }  \equiv
 \exp \left( - S_0   -
 { 1 \over 2} \ln \det \Delta \right)   \  ,   }       \\
\noalign{\vspace{1ex} }
\displaystyle{ \hspace{0ex}  {Z(T)}_{ {\rm 2-loop } }
\equiv
 \exp \left( -S_0 - { 1 \over 2} \ln \det \Delta \right)
 \times \Big[ 1+
(\mbox{2-loop Graphs})
\Big]  \ ,  }
\end{array}
\label{tree-one-two}
\end{equation}
where the 2-loop graphs are formally given by Fig.(\ref{ThirdFigure}).

Therefore the rest of the work is to fix the form of
the Green's function
eq.(\ref{green}) and find the solution ${y}_{ {}_{0} }(t)$
of the gap equation
eq.(\ref{gap}). In this paper we confine ourselves to a
time-independent
solution: those quantities obtained so far turn into the
overlined ones;
\begin{equation}
{y}_{ {}_{0} }(t) \mapsto \overline{{y}_{ {}_{0} }}:
\mbox{constant} \ ; \quad  {\rm
\Omega}(t)^2 \mapsto  \overline{{\rm \Omega}}^2   :
\mbox{constant} \
.
\end{equation}
The Green's function eq.(\ref{green}) can be obtained explicitly
\begin{equation}
\begin{array}{l}
\displaystyle{ \hspace{0ex} \overline{G}(t, t' ;
\overline{{\rm \Omega}})
\equiv  { 1
\over T}
\!
\sum_{r= -\infty}^\infty
\!  { e^{i 2 \pi r (t-t')/T}
\over
\left( { 2
\pi r \over T} \right)^2   +     \overline{{\rm \Omega}}^2 }    }
\\
\noalign{\vspace{1ex} }
\displaystyle{ \hspace{0ex} = { 1 \over 2\overline{{\rm \Omega}}
\sinh
{\overline{{\rm \Omega}} T \over 2} } \left\{  \theta(t-t')
\cosh
\overline{{\rm \Omega}} \left(  {T
\over 2} - t+t'
\right) + \theta(t'-t) \cosh \overline{{\rm \Omega}}
\left(  {T \over 2} + t- t'
\right)  \right\} \ ,  }
\end{array}
\end{equation}
where we have taken the periodic boundary condition into account.

\begin{figure}
\centering\leavevmode
\epsfysize=6cm  \epsfbox{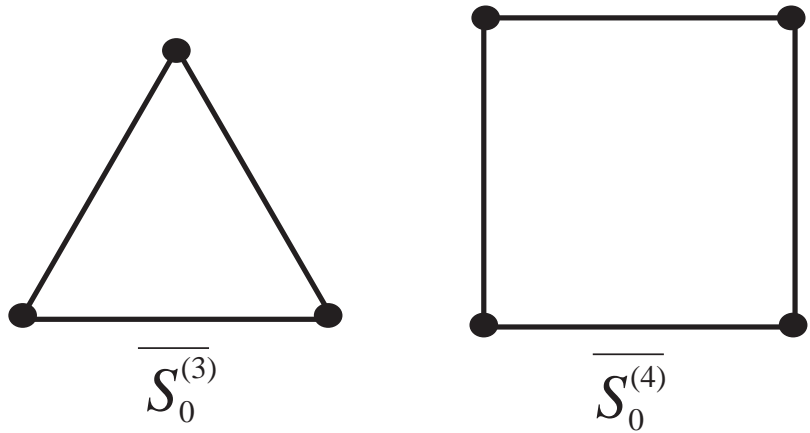}
\caption{Vertices for the constant classical solution:
the solid line represents
$\overline{G}\!\left(t, t';
\overline{\rm \Omega}\right)$ and the dot
represents the vertex which should be attached by the
double-lined $\overline{\Delta}(t, t')$.}
\label{Figure4}
\end{figure}

\begin{figure}
\centering\leavevmode
\epsfysize=9cm  \epsfbox{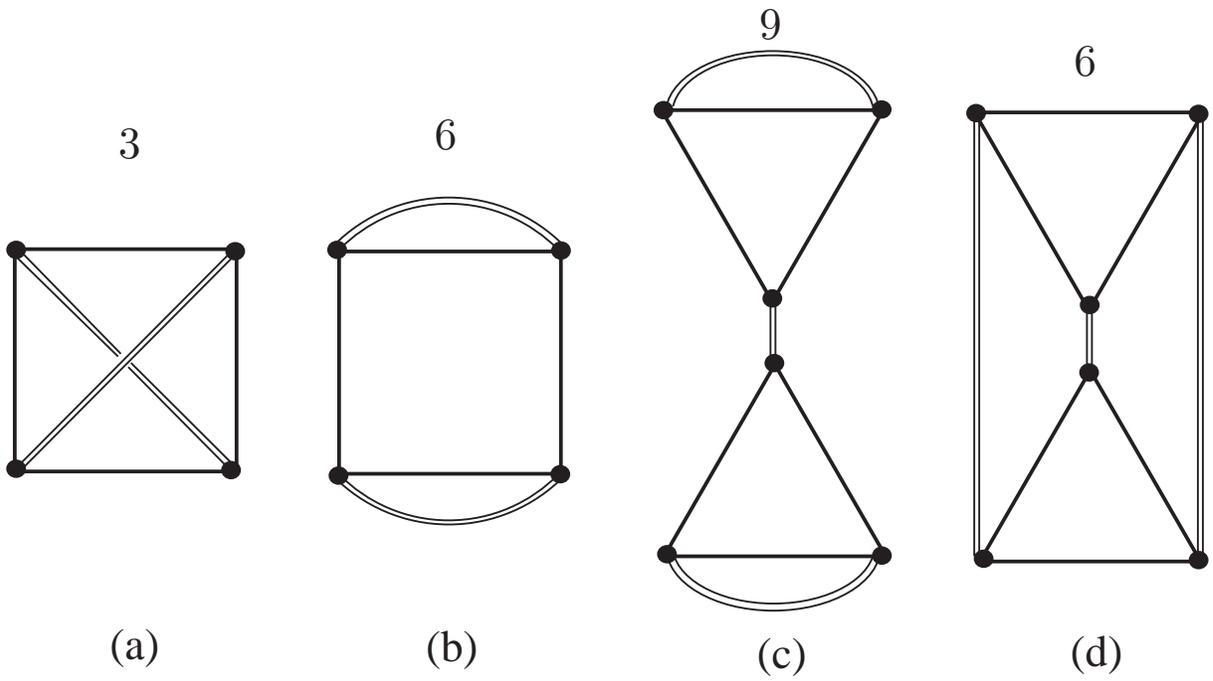}
\caption{The 2-loop Graphs for the constant classical solution:
the upper numbers in the figures represent those of multiplicity.}
\label{Figure5}
\end{figure}

Now $\overline{{\rm \Omega}}$ is the solution of the gap equation;
\begin{equation}
ig  \overline{{y}_{ {}_{0} }} = \overline{{\rm \Omega}}^2 -\omega^2
= { g^2
\over 2} \overline{G}\! \left(t, t; \overline{{\rm \Omega}} \!
\right) =  { g^2
\over 4 \overline{{\rm \Omega}} }
\coth \left( { \overline{{\rm \Omega}} T \over 2}
\right)  \ .                           \label{gap2}
\end{equation}
When $T \rightarrow$ large, $\overline{{\rm \Omega}}$ can be
expressed as
\begin{equation}
 \overline{{\rm \Omega}} ={\rm \Omega}_0 +  {\rm
\Omega}_1e^{  - {\rm \Omega}_0 T} + {\rm \Omega}_2 e^{ - 2 {\rm
\Omega}_0 T  } + \cdots \ ,
\end{equation}
where ${\rm \Omega}_0$ is the solution of the third degree equation;
\begin{equation}
{\rm \Omega}_0^3 -\omega^2{\rm \Omega}_0 = { g^2
\over 4}  \  .    \label{third}
\end{equation}
(In order to calculate the energy of the first excited state,
${\rm\Omega}_1$ must be known, which is easily obtained to be
\begin{equation}
{\rm \Omega}_1 =
{ g^2 \over 2\left(3 {\rm \Omega}_0^2 + \omega^2\right)}
\ .    
\end{equation}
In this paper, however,
only the ground state energy is considered.) Other
overlined quantities are found straightforwardly;
especially
\begin{equation}
\overline{\Delta}(t, t') =
\left( \left.{ \delta^2 S \over \delta y(t)
\delta y(t')  }\right|_{y= \overline{{y}_{ {}_{0} }}}
\right)^{-1}  =
\delta(t-t') - { g^2 \over 2 \overline{{\rm \Omega}} }
\overline{G}\! \left(t,t'; \hat{\rm \Omega}\right)  \  ,
\end{equation}
where
\begin{equation}
{\hat{\rm \Omega} }^2 \equiv 4 \overline{{\rm \Omega}}^2
+ { g^2 \over 2 \overline{{\rm \Omega}} }   \  .
\end{equation}
The tree part in eq.(\ref{tree-one-two}) becomes
\begin{equation}
{ \overline{Z}(T) }_{ {\rm Tree} }
\equiv
\exp\left( - \overline{S_0} \right)  \ ,
\end{equation}
with
\begin{equation}
\begin{array}{l}
\displaystyle{ \hspace{0ex} \overline{S_0} = T \
{ { \overline{{y}_{ {}_{0}
}}}^2 \over 2} + { 1 \over 2}
\sum_{r = - \infty}^\infty
  \ln \left\{ \left( { 2 \pi r \over T}  \right)^2   +
  {\overline{\rm \Omega}}^2   \right\}  }
  \\
\noalign{\vspace{1ex} }
\displaystyle{ \hspace{3ex}=  - { T \over 2 g^2 }
\left(  {\overline{\rm
\Omega}}^2 -
\omega^2
\right)^2 +
\ln \sinh \left( { \overline{\rm \Omega} T \over 2} \right)
+ \left(\mbox{$\overline{\rm \Omega}$-independent part} \right)
\ , }
\end{array}
\end{equation}
where the gap equation eq.(\ref{gap2}) has been utilized
to the first term in the
final expression. In the 1-loop part of eq.(\ref{tree-one-two}),
we should
know $ \ln \det
\overline{\Delta}(t, t')$ which is
\begin{equation}
\begin{array}{l}
\displaystyle{ { 1 \over 2} \ln \det \overline{\Delta}(t,
t') =  { 1 \over 2}  \sum_{r= - \infty}^\infty \ln
\left\{ 1  +  { g^2 T^2  \over 8
\pi^2 \overline{\rm
\Omega} } \
{1\over r^2 + \left(\overline{\rm \Omega}T / \pi \right)^2 }\right\} }
\\
\noalign{\vskip 1ex}
\displaystyle{ =  { 1 \over 2}
\sum_{r= - \infty}^\infty \left\{ \ln
\Big[  r^2 + \left( { \hat{\rm \Omega} T \over 2 \pi} \right)^2 \Big]
- \ln \Big[ r^2 + \left(
{ \overline{\rm \Omega} T \over \pi }  \right)^2  \Big]
\right\}   =  \ln \left(
{  \sinh \left(  \hat{\rm \Omega} T /2  \right)  \over
\sinh \left( \overline{\rm
\Omega} T \right) }  \right) \ . }
\end{array}
\end{equation}
As for the 2-loop part, the (non-local) vertices
$\overline{S_0^{(3)}}$ and $\overline{S_0^{(4)}}$ are
now expressed as in Fig.(\ref{Figure4}). Accordingly,
the 2-loop part is written as in Fig.(\ref{Figure5}):

From the graphs, we should note that there need,
in the ordinary sense, the
3- and 4-loop calculations in the 2-loop of the
auxiliary field, since our
vertices, $\overline{S_0^{(3)}}$ and
$\overline{S_0^{(4)}}$,
are non-local.
Due to this complexity, we confine ourselves
to the case
that $T \mapsto \infty$,
that is, to the ground state.
Write the Fourier transformed
$\overline{G}$ and $\overline{\Delta}$ as
\begin{equation}
\begin{array}{l}
\displaystyle{\overline{G}\!
\left(t,t';{\rm \Omega}_0 \right)
=  \int_{-\infty}^\infty { dk \over 2 \pi}
e^{-ik(t-t')}
{ 1 \over k^2 + {\rm \Omega}_0^2 }
\equiv \int_{-\infty}^\infty
{ dk \over 2 \pi} e^{-ik(t-t')} G_0(k) \ ,}
\\
\noalign{\vspace{1ex} }
\displaystyle{\overline{\Delta}(t, t') =
\int_{-\infty}^\infty { dk \over 2 \pi}
e^{-ik(t-t')}
{k^2 + 4 {\rm \Omega}^2_0
\over k^2 +{\hat{\rm \Omega}}^2 }
\equiv \int_{-\infty}^\infty
{ dk \over 2 \pi} e^{-ik(t-t')} \Delta_0(k)\  , }
\end{array}
\end{equation}
where
\begin{equation}
 {\hat{\rm \Omega}}^2
 \equiv 4 {\rm \Omega}_0^2 +
 { g^2 \over 2 {\rm \Omega}_0 }
 \ ;
\end{equation}
since $\overline{\rm \Omega}$ is
now ${\rm \Omega}_0$
under
$T \rightarrow \infty$. With these, each graph,
(a) $\sim$ (d),
can be expressed and calculated elementally as
follows:
\begin{eqnarray}
\mbox{(a)} & = &
{g^4 \over 8}
\int {d l \, dp \, dk \over (2 \pi )^3 }
\, \Delta_0(p) \Delta_0(k)
G_0(l) G_0(l+k) G_0(l+p) G_0(l+k+p)
\nonumber \\
\noalign{\vspace{-0.5ex} }
&   &   \\
\noalign{\vspace{-0.5ex} }
& = &
{g^4 \over 64 {\rm \Omega}_0^5 }
{\rho +10 \over \rho^2 (\rho +1) (\rho +2)}
\ .
\nonumber \\
\noalign{\vspace{2ex} }
\mbox{(b)} & =  &
{g^4 \over 4}
\int {d l \, dp \, dk \over (2 \pi )^3 }
\,  \Delta_0(p) \Delta_0(k)
G_0(l)^2  G_0(l+k) G_0(l+p) =
{g^4 \over 64 {\rm \Omega}_0^5 }
{  \rho +10  \over \rho^2 (\rho +1) } \ .
\\
\noalign{\vspace{2ex} }
\mbox{(c)} & =  &
-{g^6 \over 8}  \Delta_0(0)
\left[\int {d l \, dp  \over (2 \pi )^2 } \,
\Delta_0(p) G_0(l)^2 G_0(l+p) \right]^2
=  - {g^6 \over 128 {\rm \Omega}_0^8 }
{ ( \rho +6)^2 \over \rho^4 (\rho +2)^2 } \ .
\\
\noalign{\vspace{2ex} }
\mbox{(d)} & =  &
-{g^6 \over 12}
\int {d l \, dp \, dk \, dq \over(2 \pi )^4 }
\,  \Delta_0(p) \Delta_0(q) \Delta_0(p+q)
\nonumber
\\
&  &  \times  G_0(l) G_0(l+p) G_0(l+p+q)
G_0(k)  G_0(k+q) G_0(k+q+p)
\nonumber
\\
\noalign{\vspace{-0.5ex} }
&   &   \\
\noalign{\vspace{-0.5ex} }
& = & - {g^6 \over 64 {\rm \Omega}_0^8 }
{(\rho^2 + 8 \rho + 4)\over \rho^4 (\rho +1) (\rho +2)^2}
\ .
\nonumber
\end{eqnarray}
Here we have introduced a parameter,
\begin{equation}
\rho \equiv
\sqrt{{\hat{\rm \Omega}}^2 \over {\rm \Omega}_0^2 }
\  .
\end{equation}

The result for the ground state energy,
\begin{equation}
 E_0 =
 - \lim_{T \rightarrow \infty} \
 {1 \over T} \ln Z(T)  \ ,
\end{equation}
is therefore
\begin{eqnarray}
 E_0^{\rm tree} & = &
 { {\rm \Omega}_0 \over 2 }
 - { g^2 \over 32 {\rm \Omega}_0^2 }   \  ,
 \\
E_0^{1-{\rm loop} } & = & { {\rm \Omega}_0  \over 2}
\left( \rho - 1 \right)
- { g^2 \over 32
{\rm \Omega}_0^2 }  \  ,
\\
E_0^{2-{\rm loop} } & = &
{ {\rm \Omega}_0  \over 2}
\left( \rho - 1 \right)
- { g^2 \over 32 {\rm \Omega}_0^2 }
- \ { g^4 \over 64 {\rm \Omega}_0^5 }
\
{(\rho +3) (\rho +10)\over \rho^2 (\rho +1)(\rho +2) }
\nonumber
\\
\noalign{\vspace{-0.5ex} }
&   &   \\
\noalign{\vspace{-0.5ex} }
&   &  \hspace{4ex}
+ \  { g^6 \over 128 {\rm \Omega}_0^8 }  \
{\rho^3 +15\rho^2 +64 \rho+ 44 \over
\rho^4 (\rho +1)(\rho +2)^2}
\ .
\nonumber
\end{eqnarray}

Let us analyze the individual case:
\begin{itemize}
\item CASE(i); Anharmonic-Oscillator.
The solution of eq.(\ref{third}) is
given\cite{Math} by
\begin{equation}
{\rm \Omega}_0 =
\left\{ \! \!
\begin{array}{c c}
 \displaystyle{ { 2 \omega \over \sqrt{3} }
 \cos \Bigl[ { 1\over 3} \cos^{-1}
\left( { 3 \sqrt{3} g^2 \over 8 \omega^3 } \right)
\Bigr] } &
\displaystyle{: \ 0 \leq {g^2 \over 8}
\leq { \omega^3  \over 3 \sqrt{3} }  }
\\
\noalign{\vskip 2ex}
\hspace{0ex}
\displaystyle{ \sqrt[3]{ {g^2 \over 8} +
\sqrt{ {g^4 \over 64} -
{ \omega^6 \over 27}  }}  +
\sqrt[3]{  {g^2 \over 8}  -
\sqrt{ {g^4 \over 64} -
{ \omega^6 \over 27}  }}   }  &
\displaystyle{ : \  { \omega^3  \over 3 \sqrt{3} }
\leq
{ g^2 \over 8}  }
\end{array}  \right.
\end{equation}
Putting $\omega^2 \mapsto 1$ we calculate the ratio of
$E_0^{\rm
L-loop}$(L=0,1,2) to the exact numerical value
in table (c).
\end{itemize}

\vspace{1ex}

\begin{center}
{\small
\begin{tabular}{| c | c | c | c | c | }  \hline
$\displaystyle{ g^2/8 }$  &  Exact  &
$\begin{array}{c}
\mbox{Tree} \\
   \mbox{Tree/ Ex.}
\end{array}$ & $\begin{array}{c}
\mbox{1-loop} \\  \! \!
  \mbox{1-loop/ Ex.}   \! \!
\end{array}$  &
$\begin{array}{c}
\mbox{2-loop} \\
\! \!   \mbox{2-loop / Ex.}    \! \!
\end{array}$  \\   \hline
$10^{-3}$  & 0.50075 & $\begin{array}{c}
0.50025  \\   0.9990
\end{array}$  & $\begin{array}{c}
0.50075  \\   1.
\end{array}$  &  $\begin{array}{c}
0.50075  \\   1.
\end{array}$  \\   \hline
$10^{-2}$  & 0.50726 & $\begin{array}{c}
0.50248  \\   0.9906
\end{array}$   &  $\begin{array}{c}
0.50737  \\   1.0002
\end{array}$ & $\begin{array}{c}
0.50725  \\   0.9999
\end{array}$  \\  \hline
$10^{-1}$  & 0.55915 & $\begin{array}{c}
0.52290  \\   0.9352
\end{array}$   &  $\begin{array}{c}
0.56435 \\   1.009
\end{array}$ & $\begin{array}{c}
0.55775  \\   0.9975
\end{array}$  \\  \hline
$1 $  & 0.80377 & $\begin{array}{c}
  0.65268  \\   0.8038
\end{array}$ &  $\begin{array}{c}
 0.85522 \\   1.0640
\end{array}$  &  $\begin{array}{c}
 0.78548 \\   0.9772
\end{array}$ \\
\hline
$10 $  & 1.5050  & $\begin{array}{c}
1.1080    \\   0.7362
\end{array}$ &  $\begin{array}{c}
1.6729   \\   1.1116
\end{array}$  &  $\begin{array}{c}
 1.4738 \\  0.9793
\end{array}$ \\
\hline
$10^{2}$  &   3.1314 & $\begin{array}{c}
2.2356   \\   0.7139
\end{array}$ &  $\begin{array}{c}
 3.5280\\   1.1267
\end{array}$  &  $\begin{array}{c}
3.0865 \\   0.9857
\end{array}$ \\
\hline
$10^{3}$  &   6.6942 & $\begin{array}{c}
4.7445 \\   0.7088
\end{array}$ &  $\begin{array}{c}
 7.5659\\   1.1302
\end{array}$  &  $\begin{array}{c}
6.6112 \\   0.9876
\end{array}$ \\
\hline
 \end{tabular}
}
\vspace{2ex}

table(c)

\end{center}

\begin{itemize}
\item CASE(ii); Double-Well.
The solution of eq.(\ref{third}) is
\begin{equation}
{\rm \Omega}_0  =  \sqrt[3]{ \sqrt{ {g^4 \over 64} +
{ \left| \omega^2 \right|^3 \over 27}  } +
{g^2 \over 8} } -   \sqrt[3]{ \sqrt{ {g^4 \over
64} + { \left| \omega^2 \right|^3 \over 27}  }
-  {g^2 \over 8} }
\ .
\end{equation}
Putting $\omega^2 \mapsto -1$ we again compare
the result to the exact
numerical value in table (d).
\end{itemize}

\begin{center}
{\small
\begin{tabular}{| c | c | c | c | c | }  \hline
$\displaystyle{ g^2/8  }$  &  Exact  &
$\begin{array}{c}
\mbox{Tree} \\
   \mbox{Tree/ Ex.}
\end{array}$ & $\begin{array}{c}
\mbox{1-loop} \\
   \mbox{1-loop/ Ex.}
\end{array}$  &
$\begin{array}{c}
\mbox{2-loop} \\  \mbox{2-loop / Ex.}
\end{array}$  \\   \hline
 $10^{-3}$  &  -61.794 & $\begin{array}{c}
-62.500  \\   1.0114
\end{array}$  & $\begin{array}{c}
-61.794  \\   1.
\end{array}$  &  $\begin{array}{c}
-61.794   \\   1.
\end{array}$  \\   \hline
$10^{-2}$  &  -5.5532 & $\begin{array}{c}
-6.245  \\   1.1246
\end{array}$   &  $\begin{array}{c}
-5.5575  \\   1.0008
\end{array}$ & $\begin{array}{c}
-5.5541  \\   1.0002
\end{array}$  \\  \hline
$10^{-1}$  &  -0.15413  & $\begin{array}{c}
-0.57593  \\   3.7368
\end{array}$   &  $\begin{array}{c}
-0.02326  \\   0.1509
\end{array}$ & $\begin{array}{c}
-0.08479  \\   0.5501
\end{array}$  \\  \hline
$1 $  & 0.51478 & $\begin{array}{c}
  0.25  \\   0.4856
\end{array}$ &  $\begin{array}{c}
0.66421  \\   1.2903
\end{array}$  &  $\begin{array}{c}
0.41605 \\   0.8082
\end{array}$ \\
\hline
$10 $  & 1.3716  & $\begin{array}{c}
0.92366    \\   0.6734
\end{array}$ &  $\begin{array}{c}
1.5839 \\   1.1548
\end{array}$  &  $\begin{array}{c}
1.2112   \\  0.8830
\end{array}$ \\
\hline
$10^{2}$  &    3.0695 & $\begin{array}{c}
2.1501  \\   0.7005
\end{array}$ &  $\begin{array}{c}
3.4867\\   1.1359
\end{array}$  &  $\begin{array}{c}
2.7543 \\   0.8973
\end{array}$ \\
\hline
$10^{3}$  &  6.6655 & $\begin{array}{c}
4.7048   \\   0.7059
\end{array}$ &  $\begin{array}{c}
 7.5467 \\   1.1322
\end{array}$  &  $\begin{array}{c}
6.0004 \\   0.9002
\end{array}$ \\
\hline
 \end{tabular}
}
\vspace{2ex}

table(d)
\end{center}

From the above tables, in case (i) the auxiliary field
method can fit the data
within a 13\%-error under the 1-loop and a 3\%-error under the
2-loop, which is considered to be excellent. For the
Double-Well case, the
method gives us $\sim10\%$- error except the region,
$O\!\left(10^{-2}\right) < g^2<
O(1)$, where as was in the 0-dimensional case there
might need the 3-loop correction to improve the
result. Apart from this, it would be still a good
approximation for a huge coupling region.


\section{Discussion}

The auxiliary field combined with the loop-expansion can give an
excellent result for
a huge coupling region, $O\!\left(10^{-3}\right) < g^2<
O\!\left(10^3\right)$, even a component of the original variable
is single. However, in the quantum Double-Well case, there need
higher
order corrections than the 2-loop between
$O\!\left(10^{-2}\right) < g^2< O(1)$.  A maximum deviation, in
the ground state energy of 2-loop,
reaches 18 times to the exact value with the wrong sign at
$g^2 \sim 0.15$. We have calculated the first
excited energy
$E_1$ up to the 1-loop,
\begin{equation}
\Delta E \equiv E_1^{\rm 1-loop}- E_0^{\rm 1-loop} = { 2 {\rm
\Omega}_0^3 \over  3{\rm
\Omega}_0^2 + \omega^2 } \left( 1 - { g^2 \over 8  {\rm
\Omega}_0^3 } + { 3 g^2 \over 4 \sqrt{2}  {\rm
\Omega}_0^2 }  \sqrt{ 1 \over  3{\rm
\Omega}_0^2 + \omega^2 }   \right)
\end{equation}
and found a level crossing around these regions. Apparently the
approximation is broken down there. However,  it is
cumbersome to go beyond the 1-loop in quantum field theory  as well as
quantum mechanics.  The approximation scheme should be simple
and transparent.
We therefore look for another solution rather than a time-independent
solution, that is,
we must solve eq. (\ref{green}) more carefully. Indeed, the
structure of the dominant contribution to path integral has recent been
clarified by means of such as the valley method\cite{Aoyama}.
With these in
mind the work is in progress.

As for applications, the recipe is applicable almost to any situation.
Our interest is
the dynamical structure of QCD that is recently revealed in terms
of a profound
consideration into gauge invariance by Lavelle and McMullan
et al.\cite{LM,KT}, for
example. It is thus tempting to introduce this method into QCD,
which is also our future
program.

\vspace{3ex}

\centerline{\bf Acknowledgment}
\noindent
The author is grateful to Koji Harada for numerical calculations
and discussions and to Ken-Ichi
Aoki for the Double-Well numerical data.

\newpage


\end{document}